\documentclass[preprint,10pt,5p]{elsarticle}
\usepackage{amssymb,amsmath}
\usepackage{amssymb}

\usepackage{atlasphysics}
\usepackage{lineno}
\usepackage{subfigure}
\usepackage{multirow}
\usepackage{hyperref}

\begin{document}

\begin{frontmatter}

\title{Measurement of the W Charge Asymmetry in the $\Wboson \rightarrow \mu\nu$ Decay Mode in $pp$ Collisions at $\sqrt{s} = 7\TeV$ with the ATLAS detector}

\author{The ATLAS Collaboration}

\begin{abstract}

This letter reports a measurement of the muon charge asymmetry from
\Wboson~bosons produced in proton-proton collisions at a
centre-of-mass energy of $7~\TeV$ with the ATLAS experiment at the LHC. 
The asymmetry is measured in the $\Wboson \rightarrow
\mu\nu$ decay mode as a function of the muon pseudorapidity
using a data sample corresponding to a total integrated luminosity of
$31~\ipb$. The results are compared to predictions based on
next-to-leading order calculations with various parton distribution
functions. This measurement provides information on the $u$ and $d$
quark momentum fractions in the proton.

\end{abstract}

\end{frontmatter}

\section{Introduction}

The measurement of the charge asymmetry of leptons originating from
the decay of singly produced \Wboson~bosons at $pp$, $p\bar{p}$ and
$ep$ colliders provides important information about the proton
structure as described by parton distribution functions (PDFs). The
\Wboson~boson charge asymmetry is mainly sensitive to valence quark
distributions~\cite{Berger:1988tu} via the dominant production process
$u\bar{d}(\bar{u}d)\rightarrow W^{+(-)}$ and provides complementary
information to that obtained from measurements of inclusive deep
inelastic scattering cross-sections at the HERA electron-proton
collider~\cite{Chekanov:2008aa,Chekanov:2009gm,Adloff:2003uh,:2009wt}.
The HERA data do not strongly constrain the ratio between $u$ and $d$
quarks in the kinematic regime of low $x$, where $x$ is the proton
momentum fraction carried by the parton~\cite{Nakamura:2010zzi}. A
precise measurement of the \Wboson~asymmetry at the Large Hadron
Collider (LHC)~\cite{LHC:2008}, on the other hand, can contribute
significantly to the understanding of PDFs and quantum chromodynamics
(QCD) in the parton momentum fraction range $10^{-3}\lesssim x\lesssim
10^{-1}$~\cite{Alekhin:2005dx}.

In $pp$ collisions the overall production rate of \Wplus~bosons is
significantly larger than the corresponding \Wminus~rate, since the
proton contains two $u$ and one $d$ valence quarks. The first
measurements of the inclusive $\Wpm$ cross-sections at the LHC by the
ATLAS~\cite{wzpaper} and the CMS~\cite{Khachatryan:2010xn}
Collaborations confirmed the difference predicted by the Standard
Model.
The asymmetry in $pp$ collisions is symmetric with respect to the
$\Wboson$ rapidity, whereas in $\antibar{p}$ collisions it is
antisymmetric; the small sensitivity to sea-quark contributions is
strongly suppressed in $\antibar{p}$ compared to $pp$
collisions~\cite{Lohwasser:2010sp}.
Measurements in $p\bar{p}$ collisions have been performed at the
Tevatron by both the CDF~\cite{Abe:1998rv,Acosta:2005ud} and
D\O~\cite{Abazov:2007pm,Abazov:2008qv} Collaborations, also using an
iterative procedure extracting the $W$ asymmetry as a function of
$y_\mathrm{W}$ \cite{Aaltonen:2009ta}. The data have been included in
global fits of parton
distributions~\cite{Martin:2009iq,Pumplin:2002vw}.

This letter presents a differential measurement of the muon charge
asymmetry from the decay of $W^{\pm}$ bosons in $pp$ collisions at a
centre-of-mass energy of $\sqrt{s}~=~7\TeV$ at the LHC. The asymmetry
varies significantly as a function of the pseudorapidity\footnote{The
nominal $pp$ interaction point at the centre of the detector is
defined as the origin of a right-handed coordinate system. The
positive $x$-axis is defined by the direction from the interaction
point to the centre of the LHC ring, with the positive $y$-axis
pointing upwards. The azimuthal angle $\phi$ is measured around the
beam axis and the polar angle $\theta$ is the angle from the
$z$-axis. The pseudorapidity is defined as $\eta = -\ln
\tan(\theta/2)$.} $\eta_\mu$ of the charged decay lepton owing to
its strong correlation with the momentum fraction $x$ of the partons
producing the \Wboson~boson. It is defined from the cross
sections for $W \to \mu\nu$ production
$d\sigma_{\mathrm{W\mu^\pm}}/d\eta_{\mu}$ as:
\begin{equation}
 A_\mu =
  \frac{d\sigma_{\mathrm{W\mu^+}}/d\eta_\mu-d\sigma_{\mathrm{W\mu^-}}/d\eta_\mu}
       {d\sigma_{\mathrm{W\mu^+}}/d\eta_\mu+d\sigma_{\mathrm{W\mu^-}}/d\eta_\mu}\,,
  \label{eq:asymmetry}
\end{equation}
where the cross sections include the event kinematical cuts used to
select $W \to \mu\nu$ events. No extrapolation to the full phase space
is attempted in order to reduce the dependence on theoretical
predictions.

Systematic effects on the $\Wboson$-production cross-section
measurements are typically the same for positive and negative muons,
mostly canceling in the asymmetry. The ATLAS detector measures muons
with two independent detector systems.  These two
independent measurements allow systematic uncertainties to be
controlled. The results presented are based on data collected in 2010 with an
integrated luminosity of $31~\ipb$. These results significantly
improve on the previous measurement by the ATLAS
Collaboration~\cite{wzpaper}, which is based on a data set
approximately 100 times smaller.

\section{The ATLAS Detector}

The ATLAS detector~\cite{AtlasDetector,CSCbook} consists of an inner
tracking system (inner detector, or ID) surrounded by a
superconducting solenoid providing a 2T magnetic field,
electromagnetic and hadronic calorimeters and a muon spectrometer
(MS). The ID consists of pixel and silicon microstrip (SCT) detectors,
surrounded by a transition radiation tracker (TRT). The
electromagnetic calorimeter is a lead liquid-argon (LAr) detector in
the barrel and the endcap, and in the forward region copper LAr
technology is used.  Hadron calorimetry is based on two different
detector technologies, with scintillator tiles or LAr as the active
media, and with either steel, copper, or tungsten as the absorber
material. There is a poorly instrumented transition region between the
barrel and endcap calorimeter, $1.37<|\eta|<1.52$, where electrons
cannot be precisely measured. In view of a later combination, this
motivates the binning in that region for the present muon
analysis. The MS is based on three large superconducting toroids, and
a system of three stations of chambers for trigger and precise
tracking measurements.  There is a transition between the barrel and
endcap muon detectors around $|\eta|=1.05$.

\section{Data and Simulated Event Samples}\label{samplesused}

The data used in this analysis were collected from the end of
September to the end of October 2010. Basic requirements on beam,
detector, stable trigger conditions and data-quality were used in the
event selection, resulting in a total integrated luminosity of
$31~\ipb$. Events in this analysis are selected using a single-muon
trigger with a requirement on the momentum transverse to the beam
(\pT) of at least $13\GeV$. The trigger includes three levels of event
selection: a first level hardware-based selection using hit patterns
in the MS and two higher levels of software-based requirements.

Simulated event samples are used for the background estimation, the
acceptance calculation and for comparison of data with theoretical
expectations.  The processes considered are the $\Wmn$ signal, and
backgrounds from $\Wtau$, $\Zmm$, $\Ztau$, $\ttbar$ and jet production
via QCD processes (referred to as ``QCD background" in the text). The
simulated signal and background samples (except $\ttbar$) were
generated with PYTHIA 6.421~\cite{pythia} using MRST 2007
$\rm{LO}^*$~\cite{mrst} PDFs. The $\ttbar$ sample was generated with
POWHEG-HVQ v1.01 patch 4~\cite{Powheg}; the PDF set was CTEQ
6.6M~\cite{CTEQ6.6M} for the NLO matrix element calculations, while
CTEQ 6L1 was used for the parton showering and underlying event via
the POWHEG interface to PYTHIA. The radiation of photons from charged
leptons was treated using PHOTOS v2.15.4~\cite{Photos} and TAUOLA
v1.0.2~\cite{Tauola} was used for tau decays. The underlying and
pile-up events were simulated according to the ATLAS MC09
tune~\cite{MC09tune}.  The generated samples were passed through the
GEANT4\\\cite{geant4} simulation of the ATLAS
detector~\cite{AtlasSimulation}, reconstructed and analysed with the
same analysis chain as the data. The cross-section predictions for $W$
and $Z$ were calculated to next-to-next-to-leading-order (NNLO) using
FEWZ~\cite{FEWZ} with the MSTW 2008~\cite{MSTW2008} PDFs. The $\ttbar$
cross-section was obtained at next-to-leading-order (plus
next-to-next-to-leading-log, NNLL) using
POWHEG~\cite{Bonciani:1998vc}. The Monte Carlo (MC) were generated
with, on average, two soft inelastic collisions overlaid on top of the
hard-scattering event.  Events in the MC samples were weighted so that
the distribution of the number of inelastic collisions per bunch
crossing matched that in data, which has an average of $2.2$.

\section{Event Selection}

The criteria for the event selection and muon identification follow
closely those used for the $W$ boson inclusive cross-section
measurement~\cite{wzpaper}, with an improved muon quality
selection~\cite{wjets}.  Events from $pp$ collisions are selected by
requiring a collision vertex with at least three tracks each with
transverse momentum greater than $150\MeV$. A beam-spot constraint has
been applied in the collision vertex reconstruction stage
significantly improving the resolution on the collision vertex
position in the transverse plane. To reduce the contribution of
cosmic-ray and beam-halo events, induced by proton losses from the
beam, the analysis requires the collision vertex position along the
beam axis to be within~$20\cm$ of the nominal interaction point. The
collision vertex is defined as the vertex closest in $z$ to the
selected muon.

Events with a high transverse momentum muon are selected by imposing
stringent requirements to ensure good discrimination of $\Wmn$ events
from background. The muon parameters are first reconstructed
separately in the MS and ID. Subsequently, the tracks from the ID and
MS are matched. Their parameters are then combined, weighted by their
respective errors, to form a combined muon. The \Wboson\ candidate
events are required to have at least one combined muon track with
$\pT>20\GeV$ and {\pT} measured by the MS alone greater than
$\pT^{\mathrm{MS}}>10\GeV$, within the range $|\eta_{\mu}|<2.4$. The
difference between the ID and MS \pT, corrected for the mean energy
loss in the material traversed between the ID and MS, is required to
be less than 0.5 times the ID \pT,

$$
\pT^{\mathrm{MS}} ({\mathrm{energy\;loss\;corrected}})- \pT^{ID}<0.5\phantom{0}\pT^{\mathrm{ID}}.
$$

This requirement increases the robustness against track reconstruction
mismatches, including decays-in-flight of\\ hadrons. In addition, a
minimum number of hits in the ID is required to ensure high quality
tracks~\cite{wjets}. In order to further reduce non-collision
backgrounds, the difference between the $z$ position of the muon track
extrapolated to the beam line and the $z$ coordinate of the collision
vertex is required to be less than $1\cm$.  A track-based isolation
for the muon is defined as $\sum \pT^{\mathrm{ID}}/\pT<0.2$, where
$\sum\pT^{\mathrm{ID}}$ is the scalar sum of transverse momenta of all
other tracks measured in the ID belonging to the same collision vertex
within a cone\footnote{$\Delta R$ is defined as $\Delta R=\sqrt{\Delta
\eta^2 + \Delta \phi^2}$.}  $\Delta R <0.4$ around the muon direction
excluding the ID track associated with the muon, and $\pT$ is the
transverse momentum of the muon combined track.

The reconstruction of the missing transverse energy ($\met$) and the
transverse mass (\mT) follows the prescription in~\cite{wzpaper}. The
\met~is determined from the energy deposits of calibrated calorimeter
cells in three-dimensional clusters and is corrected for the momentum
of all muons reconstructed in the event. Jet-quality requirements are
applied to remove a small fraction of events where sporadic
calorimeter noise and non-collision backgrounds can affect the
\met~reconstruction~\cite{CLEANING}. The transverse mass is defined
as

\begin{equation}
\ensuremath{\mT} =
\sqrt{2\ensuremath{\pT}^{\mu}\ensuremath{\pT}^{\nu}(1-\cos(\phi^{\mu}-\phi^{\nu}))},
\end{equation}

where the highest $\pt$ muon is used and the $(x,y)$ components 
of the neutrino momentum are inferred
from the corresponding \met~components. Events are required to
have $\met>25\GeV$ and $\mT>40\GeV$, yielding 129157 
$\Wboson$ candidates.

\section{$W^{\pm}$ Signal Yield and Background Estimation}
\label{sec:background}

\begin{figure*}[t]
\centering
\subfigure[]{\includegraphics[width=0.40\linewidth]{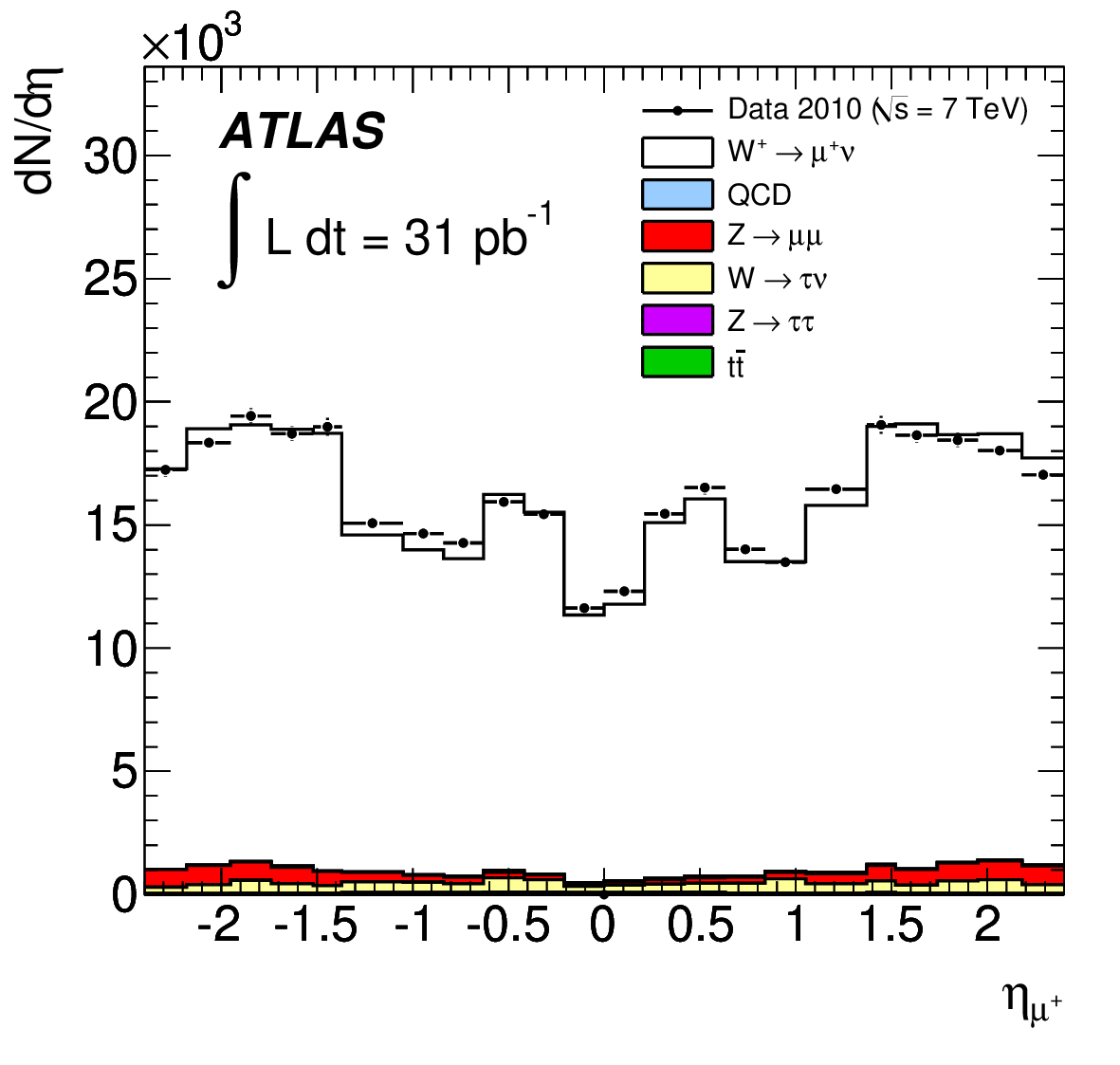}\label{subfig:etapos}}
\subfigure[]{\includegraphics[width=0.40\linewidth]{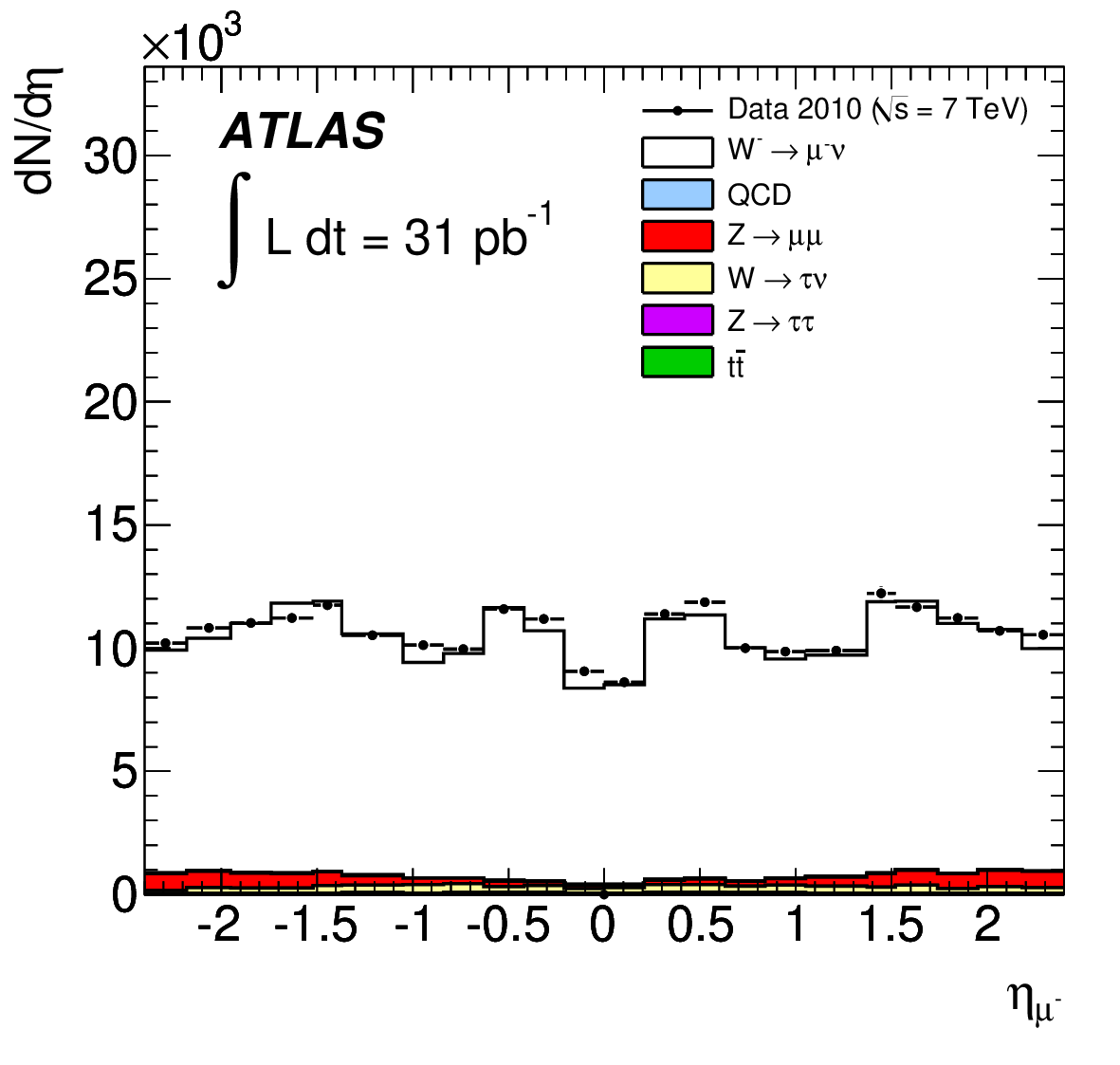}\label{subfig:etaneg}}
\caption{Distribution of the muon pseudorapidity $\eta_\mu$ of
  $\Wplus$~\subref{subfig:etapos} and $\Wminus$~\subref{subfig:etaneg}
  candidates, after final selection. The data are compared to PYTHIA
  MC simulation, broken down into the signal and various background
  components. POWHEG is used for the \ttbar\ background. The MC
  distributions are normalised to the total number of events in data.}
\label{fig:eta}
\end{figure*}

\begin{table*}[ht]
\begin{center} 

\begin{tabular}{l|ccc|ccc}
\hline
\hline 
         &   \multicolumn{3}{|c|}{$\mu^+$} &     \multicolumn{3}{c}{$\mu^-$} \\\cline{2-7}
         & Observed  &  Exp. Background & $\CWmp$  & Observed  &  Exp. Background & $\CWmm$  \\
\hline 
$0.00<|\eta_\mu|<0.21$  & $5028$ & $272\pm 51$ & $0.594\pm 0.005$ & $3711$ & $236\pm 55$ & $0.584\pm 0.004$\\
$0.21<|\eta_\mu|<0.42$  & $6486$ & $385\pm 70$ & $0.779\pm 0.009$ & $4736$ & $334\pm 70$ & $0.759\pm 0.008$\\
$0.42<|\eta_\mu|<0.63$  & $6818$ & $481\pm 88$ & $0.808\pm 0.009$ & $4923$ & $357\pm 70$ & $0.800\pm 0.009$\\
$0.63<|\eta_\mu|<0.84$  & $5939$ & $366\pm 76$ & $0.686\pm 0.008$ & $4194$ & $329\pm 64$ & $0.691\pm 0.008$\\
$0.84<|\eta_\mu|<1.05$  & $5909$ & $395\pm 63$ & $0.672\pm 0.007$ & $4195$ & $358\pm 63$ & $0.681\pm 0.008$\\
$1.05<|\eta_\mu|<1.37$  &$10086$ & $627\pm 93$ & $0.735\pm 0.007$ & $6531$ & $585\pm 101$ & $0.752\pm 0.007$\\
$1.37<|\eta_\mu|<1.52$  & $5708$ & $363\pm 57$ & $0.905\pm 0.009$ & $3595$ & $348\pm 59$ & $0.914\pm 0.009$\\
$1.52<|\eta_\mu|<1.74$  & $8218$ & $542\pm 89$ & $0.905\pm 0.008$ & $5035$ & $518\pm 82$ & $0.925\pm 0.008$\\
$1.74<|\eta_\mu|<1.95$  & $7956$ & $605\pm 114$ & $0.896\pm 0.009$ & $4671$ & $456\pm 80$ & $0.898\pm 0.008$\\
$1.95<|\eta_\mu|<2.18$  & $8364$ & $647\pm 100$ & $0.903\pm 0.009$ & $4952$ & $548\pm 91$ & $0.910\pm 0.009$\\
$2.18<|\eta_\mu|<2.40$  & $7541$ & $534\pm 81$ & $0.881\pm 0.010$ & $4561$ & $492\pm 82$ & $0.896\pm 0.010$\\
\hline 
\hline 
\end{tabular}
\caption{Summary of observed number of events, expected background and
  correction factor $\CWm$ for positive and negative muons in bins of
  $|\eta_{\mu}|$. The errors given for the background estimates
  include systematic uncertainties, including the uncertainty due to
  the luminosity, used in the normalization of the electro-weak and
  $\ttbar$ components. The $\CWm$ factors include trigger and muon
  reconstruction scale factors; they include the statistical
  uncertainty from the MC sample and the trigger and reconstruction
  scale factors.  }
\label{tab:Wsummary_muon}
\end{center} 
\end{table*} 

Many components in the $W$ cross-section measurement, such as the
luminosity or detector efficiencies, are in principle the same for
positive and negative muons and therefore mostly cancel in the
asymmetry calculation.  The main experimental biases on the asymmetry
measurement come from possible differences in the reconstruction of
positive and negative muons.  Each effect (trigger and reconstruction
efficiency and momentum scale) is examined to check that the two
charges behave in the same way within the systematic uncertainties.
These studies are performed in absolute pseudorapidity in order to
reduce the uncertainty associated with the limited size of the data
samples used.

As in past $W$ analyses, trigger~\cite{wjets} and muon
reconstruction~\cite{wzpaper,wjets} efficiencies as a function of muon
$\eta_\mu$ have been measured in data using a sample of unbiased muons
from \Zmm~decays, which provides a source of muons with small
background.  The trigger efficiency is determined relative to a
reconstructed muon satisfying the selection criteria of the
analysis. The average trigger efficiencies after the full $\Wboson$
selection are $(81\pm 2)$\% in the central detector region or
low-$\eta$ region, $|\eta_\mu|<1.05$, and $(94\pm 1)$\% in the forward
detector region or high-$\eta$ region, $1.05<|\eta_\mu|<2.4$, where
the differences are due to the geometrical acceptance of the muon
trigger chambers.  In the same muon sample, the muon reconstruction
efficiency relative to an ID track is measured to be $(93\pm 1)$\%
overall. The efficiency for reconstructing an ID track is $(99\pm
1)$\%~\cite{wzpaper}. The quoted uncertainties on these efficiencies
are statistical.

Corrections have been applied to the simulated samples to account for
differences in the trigger and reconstruction efficiencies between
data and simulation. These are based on the ratio of the efficiency in
data and in simulation, and are computed as a function of the muon
$\eta_\mu$ and charge. The corrections for each charge agree within
the statistical uncertainties, so the charge-averaged result is
applied. For the trigger, the corrections are $0.98$ and $1.03$ in the
central and forward MS regions, respectively. For the reconstruction
efficiency, the correction factors are about $0.99$ per $\eta_\mu$ bin
except for the central-forward MS transition region ($|\eta_\mu|$
about $1.05$) where the correction factor is $0.94$.

The muon momentum resolution is affected by the a\-mount of material
traversed by the muon, the spatial resolution of the individual track
points and the degree of internal alignment of the ID and
MS~\cite{muperfcosmics}. This resolution has been measured as a
function of $\eta_\mu$ for the main detector regions 
(in $\eta_\mu$ ranges delimited by $1.05, 1.7, 2.0$ and $2.4$) from the
width of the di-muon invariant mass distribution in $\Zmm$ decays and
from the comparison of the momentum measurements in the ID and MS in
$\Zmm$ and $\Wmn$ decays. The measured
resolution is worse than expected from simulation by
$1\;\!$--$\;\!5\%$, with the maximum discrepancy reached in the high-$\eta_\mu$
region of the detector~\cite{muonreso}. 
The discrepancy is due to residual mis-alignments in the ID and MS, 
imperfections in the description of the inert material in simulation and
an imperfect mapping of the magnetic field in the MS transition region where
the field is highly non-uniform. Smearing corrections are therefore applied to
the simulation in order to improve the agreement with data.

If the accuracy of the muon momentum measurement is different for
positive and negative muons, this difference can produce a bias in the
acceptance of $\mu^+$ with respect to $\mu^-$. Differences in the muon
\pT\ measurement between data and simulation have been evaluated
comparing the curvature of muons from \Wboson\ candidates in data and
in templates derived from simulation. A binned likelihood fit for a
momentum-scale correction that yields the best agreement between data
and simulation is performed as a function of $\eta_\mu$ separately for
positive and negative charges. The measured biases in the $\pT$ scale
between the two charges are $<1\%$, but they increase to about
$3\%$ in the transition and high-$\eta_\mu$ regions due to residual
mis-alignments in the ID and MS. These corrections are applied to the
muon momenta in the simulated samples.

Figure~\ref{fig:eta} shows the
pseudorapidity distribution of the selected positive and negative
muons. Data distributions are compared to the PYTHIA MC simulation,
normalised to the total number of events in data. The shape of the
simulation agrees well with the shape of the data after the
corrections for the reconstruction and trigger efficiencies, and the
muon-momentum scale and resolution.

The main backgrounds to $\Wmn$ arise from heavy flavour decays in
multijet events and from the electro-weak background from \Wtau~where
the tau decays to a muon, \Zmm~where one muon is not reconstructed and
produces fake \MET, and \Ztau~where one tau decays to a muon, as well
as semileptonic \ttbar~decays in the muon channel. Backgrounds from
the production of di-bosons ($WW$, $WZ$ and $ZZ$) and single top
quarks are found to be negligible. The \Wtau~contribution is treated as a
background. While this contribution presents the same asymmetry as the
\Wmn~signal, it is difficult to include in PDF fits, which assume that
the asymmetry is a function of $\eta_\ell$ for $\Wboson \rightarrow
l\nu$. No explicit veto on events with a second muon is applied.

The background estimates of the electro-weak and $\ttbar$ backgrounds
and the QCD background closely follow the methods used in the $W$
inclusive cross-section measurement~\cite{wzpaper}. They are
determined separately for positive and negative muons as a function of
$\eta_{\mu}$.  The electro-weak and $\ttbar$ backgrounds are estimated
using MC simulation.  The QCD background comes primarily from $b$ and
$c$ quark decays, with a smaller contribution from pion and kaon
decays in flight. This background is estimated using a data-driven
method similar to the one described in~\cite{wzpaper}. The sample of
events fulfilling the full $\Wboson$ selection criteria with the
exception of the muon isolation requirement is compared before and
after the isolation requirement. The isolation efficiency for non-QCD
events is measured in data with the $\Zmm$ sample. The efficiency for
QCD events is estimated in a control sample of low-$\pt$ muons
extrapolated to the high-$\pt$ and high-$\met$ signal region using the
simulated jet sample. Since the samples before and after isolation can
be defined in terms of a QCD and non-QCD component, the expected
number of QCD events can thus be determined.

\begin{figure*}[tph]
\centering
\subfigure[]{\includegraphics[width=0.4\linewidth]{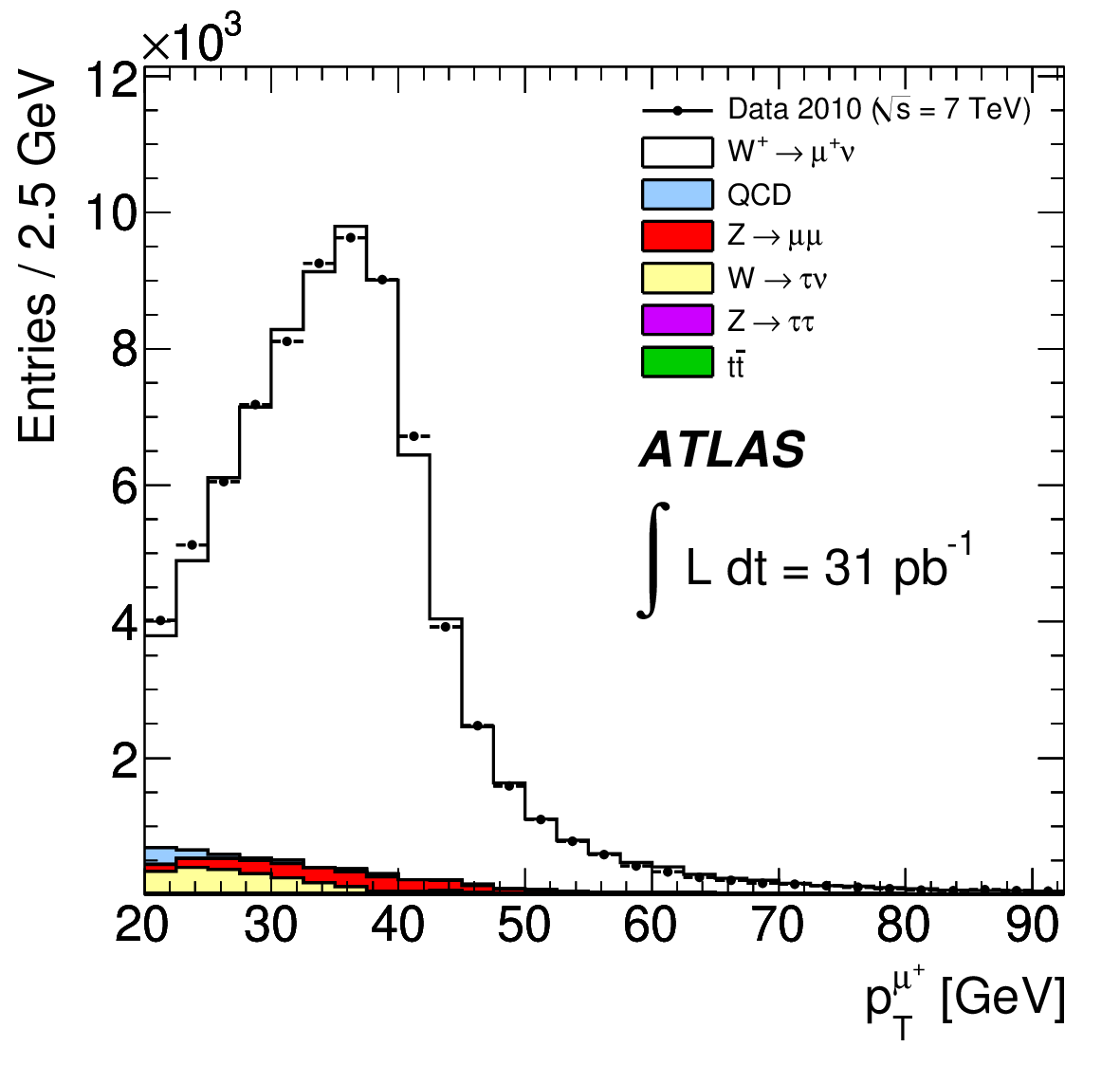}\label{subfig:muptpos}}
\subfigure[]{\includegraphics[width=0.4\linewidth]{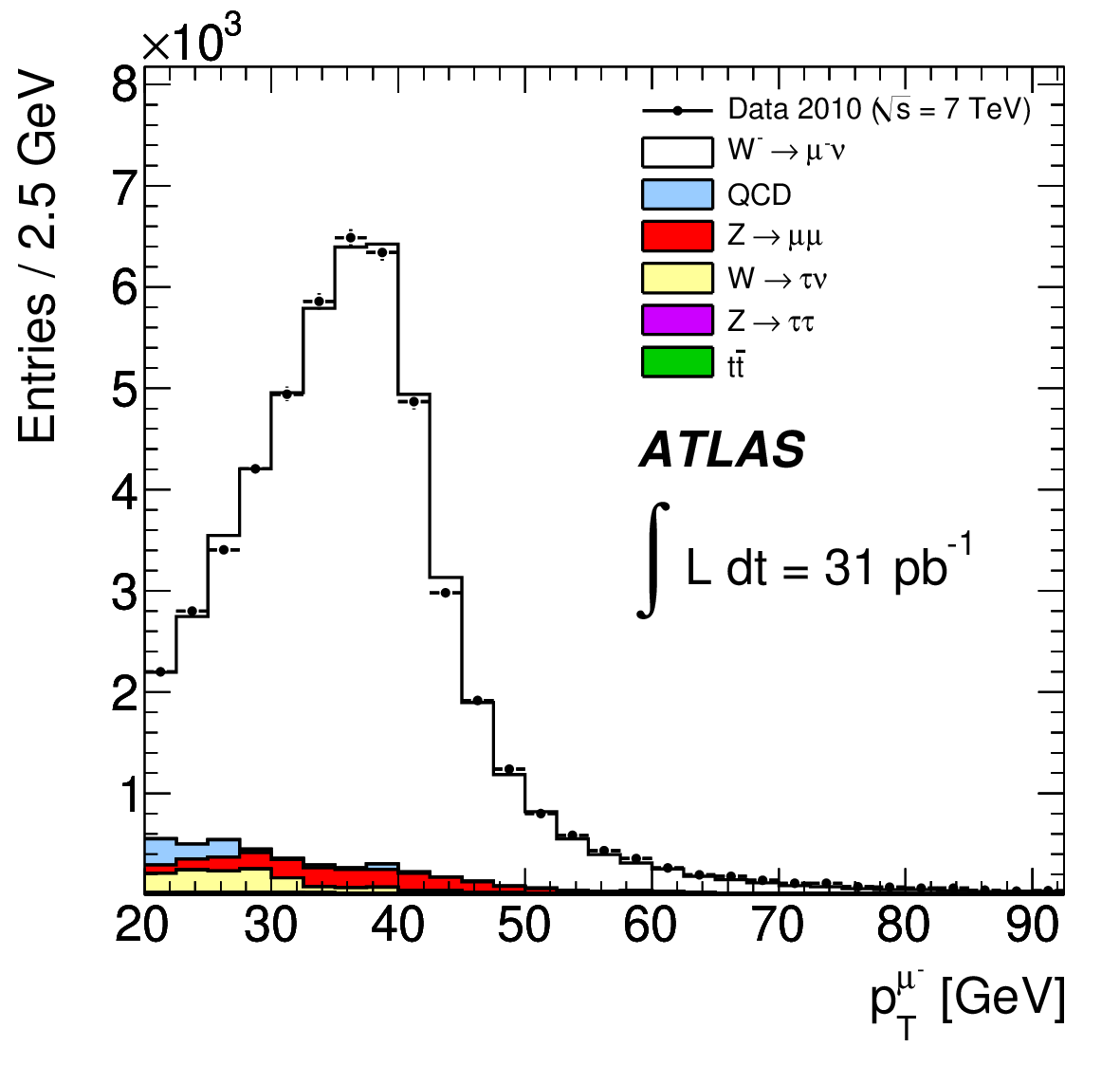}\label{subfig:muptneg}}
\caption{Distribution of the transverse momentum of positive and
  negative muons after the final selection. The data are compared to
  PYTHIA MC simulation, broken down into the signal and various
  background components. POWHEG is used for the \ttbar\
  background. The MC distributions are normalised to the total number
  of entries in data.}
\label{fig:mupt}
\end{figure*}
  
The expected background amounts to $7\%$ of the selected events; 
$6\%$ is the electro-weak and $\ttbar$ contribution ($3\%$ $\Zmm$,
$2\%$ $\Wtau$, and $1\%$ for the sum of $\ttbar$ and $\Ztau$) and the
remainder is the QCD background. The cosmic ray
background contamination is estimated to be smaller by a factor of $10^5$ 
compared to the signal and thus negligible.  The $W^{\pm}$ candidate events 
and expected background contributions are summarised in 
Table~\ref{tab:Wsummary_muon}.

Figure~\ref{fig:mupt} shows the transverse momentum distribution for
positive and negative muons after the full event selection. They are
compared with the distributions predicted by the corrected PYTHIA MC
simulation normalised to the total number of events in data.
The discrepancies between data and MC are taken into account by
the systematic uncertainty assigned to the measurement of muon momenta
explained in Section~\ref{sec:syst}.
The correction factors, $\CWm$, corresponding to the ratio of
reconstructed over generated events in the simulated $W$ sample,
satisfying all kinematic requirements of the event selection,
$p_{\mathrm{T}}^{\mu} > 20\GeV$, $p_{\mathrm{T}}^{\nu} > 25\GeV$, $\mT
> 40\GeV$, are also listed in Table~\ref{tab:Wsummary_muon}. No
correction is made to the full acceptance.  The $\CWm$ factors include
trigger and muon reconstruction scale factors to correct for observed
deviations between data and MC efficiencies. Due to a reduced
geometric acceptance in the trigger, the $\CWm$ factors for the lowest
$|\eta_\mu|$ bins are significantly smaller than those for the higher
$|\eta_\mu|$ regions.

\section{Systematic Uncertainties}
\label{sec:syst}

All systematic uncertainties on the asymmetry measurement are
determined in each $|\eta_\mu|$ bin accounting for correlations
between the charges and are summarised in
Table~\ref{tab:muonsysts_summary_emu_abs}. The dominant sources of
systematic uncertainty on the asymmetry come from the trigger and
reconstruction efficiencies. The determination of these efficiencies
is affected by the statistical uncertainty due to the small available
sample of $\Zmm$ events.  Systematic uncertainties on the efficiencies
are determined from studies of the impact of the selection criteria
and backgrounds, and no significant charge biases are found.  There is
a loss of trigger efficiency in the low pseudorapidity region due to
reduced geometric acceptance, resulting in a larger statistical
error. As a result, the trigger systematic uncertainty on the
asymmetry is largest in the low pseudorapidity bins (6-7\% for central
$|\eta_\mu|$ and 2-3\% for forward $|\eta_\mu|$).  Similarly, the
uncertainties associated with the reconstruction efficiency are larger
in the lowest pseudorapidity bin (about $7\%$), and in the MS
central-forward transition region (about $3\%$), due to geometrical
acceptance effects associated with reduced chamber coverage. In the
remaining regions, the uncertainty is about 1-2\%.

\begin{table*}[hb]
{\small
  \centering
  \begin{tabular}{l|ccccccc}
\hline
\hline
& \multirow{2}{*}{Trigger} & \multirow{2}{*}{Reconstruction} & $\pt$ Scale and & QCD  & Electro-weak and $\ttbar$ & Theoretical\\
  & & & Resolution & Normalisation & Normalisation & Modelling\\
\hline
$0.00<|\eta_{\mu}|<0.21$ & $0.011$ & $0.010$ & $0.003$ & $0.003$ & $<0.001$ & $0.007$\\
$0.21<|\eta_{\mu}|<0.42$ & $0.010$ & $0.004$ & $0.003$ & $0.003$ & $<0.001$ & $0.005$\\
$0.42<|\eta_{\mu}|<0.63$ & $0.009$ & $0.004$ & $0.003$ & $0.003$ & $<0.001$ & $0.006$\\
$0.63<|\eta_{\mu}|<0.84$ & $0.012$ & $0.004$ & $0.003$ & $0.002$ & $0.001$ & $0.007$\\
$0.84<|\eta_{\mu}|<1.05$ & $0.013$ & $0.006$ & $0.003$ & $0.003$ & $0.001$ & $0.008$\\
$1.05<|\eta_{\mu}|<1.37$ & $0.006$ & $0.007$ & $0.002$ & $0.002$ & $0.001$ & $0.006$\\
$1.37<|\eta_{\mu}|<1.52$ & $0.006$ & $0.005$ & $0.002$ & $0.003$ & $0.002$ & $0.005$\\
$1.52<|\eta_{\mu}|<1.74$ & $0.005$ & $0.004$ & $0.002$ & $0.003$ & $0.002$ & $0.007$\\
$1.74<|\eta_{\mu}|<1.95$ & $0.006$ & $0.003$ & $0.002$ & $0.002$ & $0.001$ & $0.006$\\
$1.95<|\eta_{\mu}|<2.18$ & $0.006$ & $0.004$ & $0.002$ & $0.003$ & $0.002$ & $0.009$\\
$2.18<|\eta_{\mu}|<2.40$ & $0.007$ & $0.005$ & $0.002$ & $0.003$ & $0.002$ & $0.007$\\
\hline
\hline
  \end{tabular}
  \caption{Absolute systematic uncertainties on the $\Wboson$ charge
    asymmetry from different sources as a function of absolute muon pseudorapidity 
    that are described in the text.}
    \label{tab:muonsysts_summary_emu_abs}
} 
\end{table*}

The muon momentum scale and resolution corrections contribute to the
uncertainty primarily due to the limited statistics for the fitting
procedures used to measure the differences between the data and
simulation. An additional source of uncertainty arises from potential
biases in the template shapes. The size of this effect is determined
by using different templates created by shifting the resolution
parameters in opposite directions to account for possible charge
biases.  Uncertainties associated with the modelling of the background
contributions to the templates, particularly the QCD background, are
also included. The resulting uncertainty on the asymmetry is in the
1-2\% range, with little dependence on $\eta_\mu$.  The redundant ID
and MS momentum measurements result in a rate of charge
mis-identification smaller than $10^{-4}$ in the $\pt$ range
considered, resulting in a negligible impact on the asymmetry.

\begin{figure}[tbph]
\centering
\includegraphics[width=0.80\linewidth]{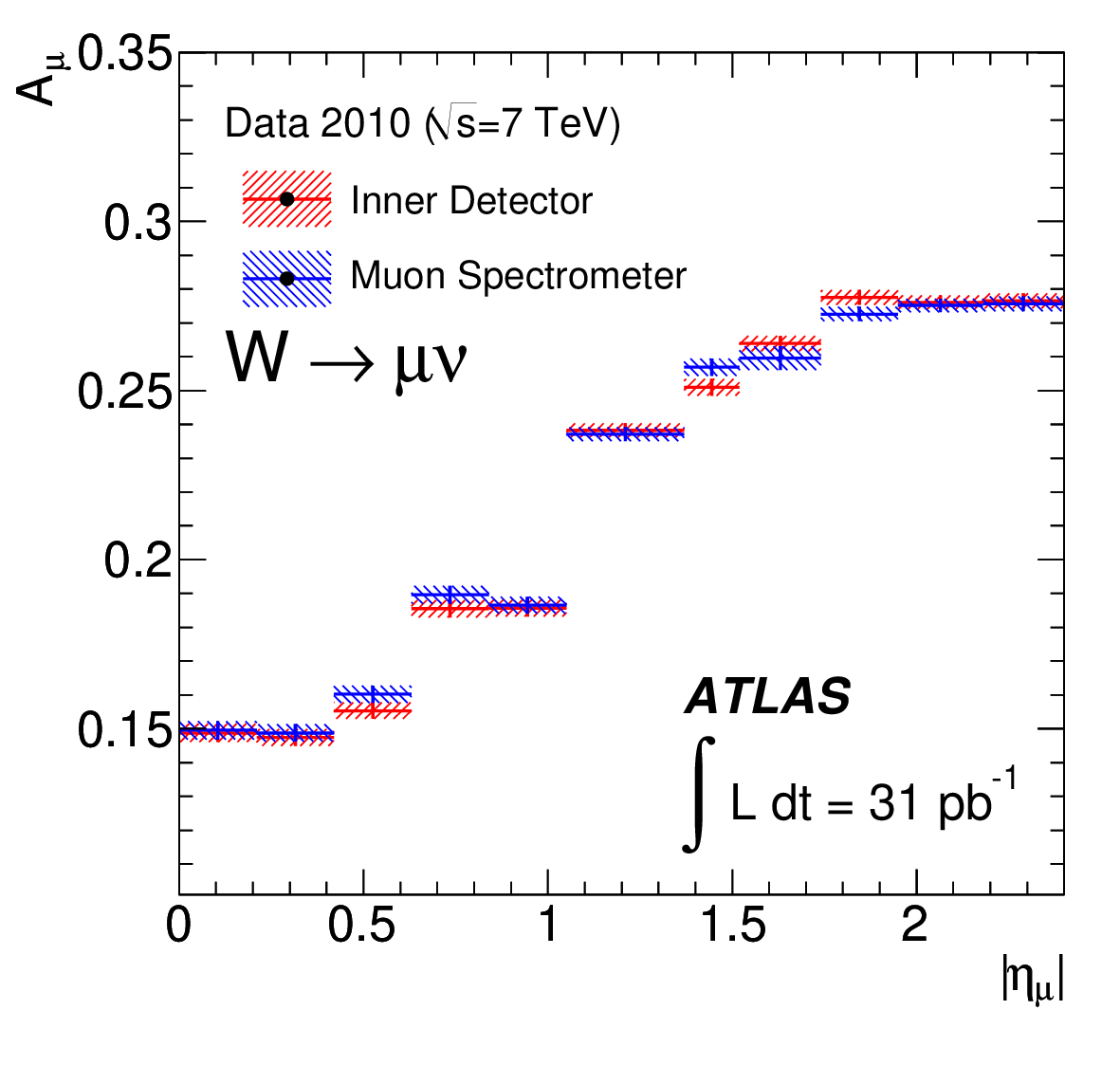}
\caption{\Wboson\ charge asymmetry measured using the ID and MS
  separately. The MS measurement is extrapolated to
  the collision vertex, and corrected for energy-loss in the
  calorimeters. The two measurements are independently corrected for
  effects of the muon-momentum scale on the muon acceptance.  The two
  measurements are statistically correlated to a large extent, since they use
  the same muons reconstructed by different subdetectors and algorithms. 
  The error bar reports therefore only the systematic uncertainty associated 
  with the momentum-scale
  correction.}
\label{fig:idmsasym}
\end{figure}

The momentum-scale correction procedure is further tested by exploiting the
redundant muon-momentum measurements offered by the ATLAS
detector. The full asymmetry measurement is performed with the ID and
MS components of the combined muon separately, including the scale
corrections. Figure~\ref{fig:idmsasym} compares the two independently
corrected charge-asymmetry distributions, showing good agreement 
within the systematic uncertainty associated with the momentum-scale
correction. 

The systematic uncertainties on the QCD background arise primarily
from the uncertainty on the isolation efficiency for muons in QCD
events due to possible mis-modellings of the extrapolation of the
isolation efficiency to the large $\pt$ and $\MET$ region in the QCD
simulation (40\%).  This has been derived from differences in the
efficiency predictions between data and simulation in the low muon
$\pt$ control region and in sideband regions where the muon $\pt$ or
$\MET$ cuts are reversed. The electro-weak and $\ttbar$ background and
signal contributions are subtracted from data in these
comparisons. Additional uncertainties due to the non-QCD isolation
efficiency and the statistical uncertainty are included in the total
uncertainty on the QCD background estimate.  The corresponding
systematic uncertainty on the asymmetry is 1-2\%, with little
dependence on $\eta_\mu$.

For the electro-weak and $\ttbar$ backgrounds, the uncertainties in the
cross-sections include the PDF uncertainties (3\%), and the
uncertainties estimated from varying the renormalization and
factorization scales: 5\% for $W$ and $Z$, and 6\% for
$\ttbar$~\cite{Moch:2008qy,Beneke:2009ye,wzpaper}. These uncertainties
are taken as overall uncertainties, no $\eta$ dependence is
accounted for. An additional uncertainty from the luminosity of $11\%$
is included, since the backgrounds are scaled to the luminosity
measured in data. The combination of all these contributions results
in an uncertainty on the asymmetry of less than $1\%$.

The impact of using an NLO MC using the CTEQ 6.6 PDF rather than
PYTHIA with MRST LO* PDF in the $\CWm$ factor calculation has been
evaluated and an additional systematic uncertainty of about $3\%$ is
included to account for the small variations observed, as listed in
Table~\ref{tab:muonsysts_summary_emu_abs} as the uncertainty due to
theoretical modelling.  PYTHIA uses a leading-log calculation for
$\Wboson$ production and is expected to give a reasonably accurate
prediction for the low $\Wboson$ transverse momentum $\Wpt$ region
whereas MC@NLO~\cite{Frixione:2002ik} uses higher-order matrix
elements and is therefore expected to be more reliable in the high
$\Wpt$ region. Therefore the differences in the scale factors
associated with these two MC calculations gives a reasonable estimate
of the associated systematic error.

\section{Results and Conclusions}

\begin{figure}[tbph]
\centering
\includegraphics[width=0.95\linewidth]{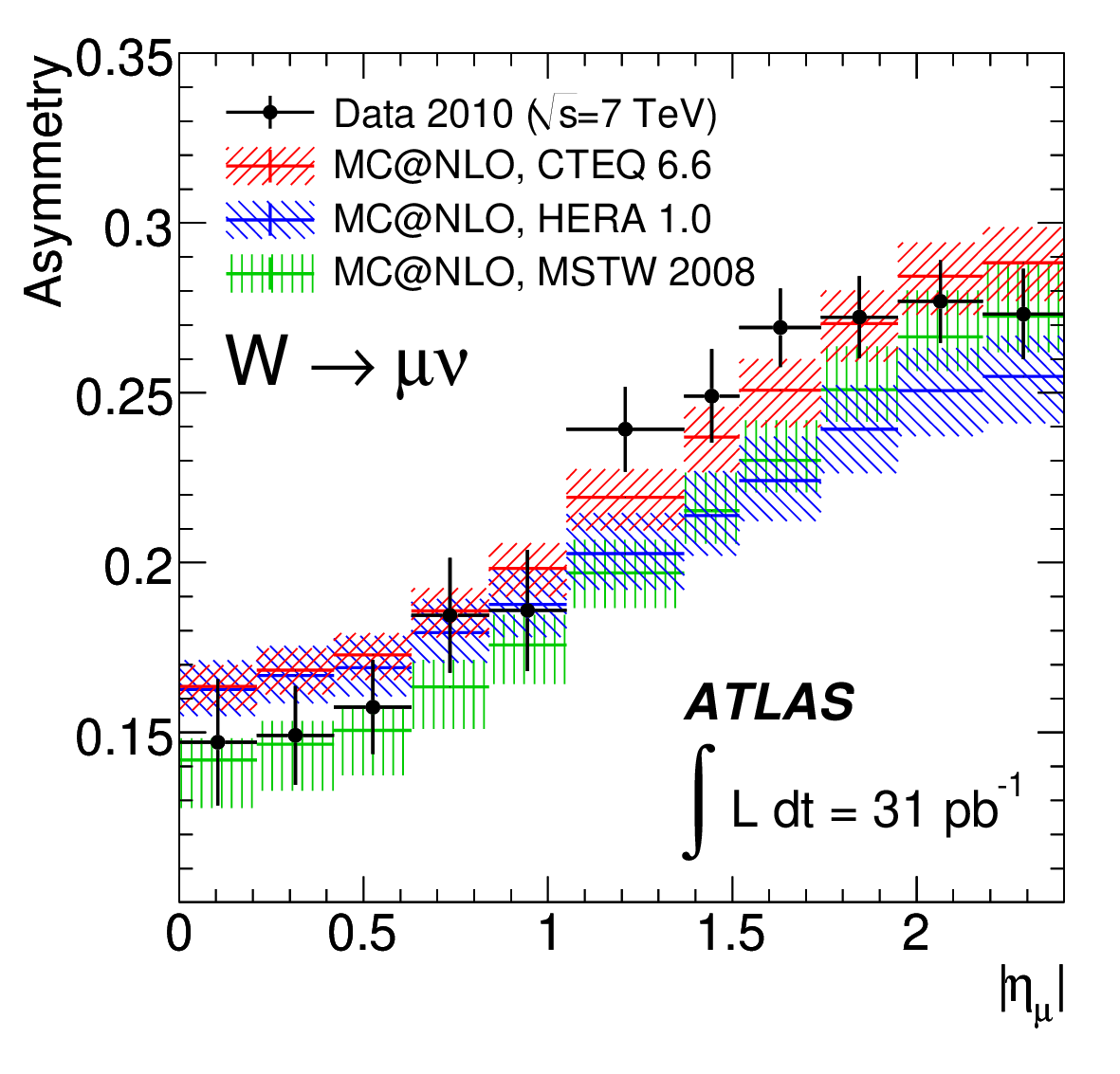}
\caption{The muon charge asymmetry from \Wboson-boson decays in bins
  of absolute pseudorapidity. 
  The kinematic requirements applied are
  $p_{\mathrm{T}}^{\mu} > 20\GeV$, $p_{\mathrm{T}}^{\nu} > 25\GeV$ 
  and $\mT > 40\GeV$.
  The data points (shown with error
  bars including the statistical and systematic
  uncertainties) are compared to MC@NLO predictions with different PDF sets.
  The PDF uncertainty bands are described in the text and include
  experimental uncertainties as well as model and parametrization
  uncertainties.}
\label{fig:asym}
\end{figure}

The measured differential muon charge asymmetry in eleven bins of muon
absolute pseudorapidity is shown in Table~\ref{tab:asymvals} and
Figure~\ref{fig:asym}. The statistical and systematic uncertainties
per $|\eta_\mu|$ bin are included and contribute comparably to the
total uncertainty. Table~\ref{tab:asymvals} and Figure~\ref{fig:asym}
also show expectations for the muon asymmetry from $W$ predictions at
NLO with different PDF sets: CTEQ 6.6~\cite{Pumplin:2002vw}, HERA
1.0~\cite{:2009wt} and MSTW 2008~\cite{Martin:2009iq}; all predictions
are presented with 90\% confidence-level error bands. All MC
predictions are calculated using MC@NLO, with all kinematic selection
criteria applied to the truth particles. The PDF uncertainty bands are
obtained by summing in quadrature the deviations of each of the PDF
error sets~\cite{Pumplin:2001ct} from the respective nominal
predictions, according to the specifications of the corresponding PDF
collaborations to get 90\% C.L. bands. These uncertainties for all
predictions include experimental uncertainties as well as model and
parametrization uncertainties. The HERA 1.0~\cite{:2009wt} set also
includes the uncertainty in $\alpha_s$, which, however, is not the
dominant source of uncertainty.

While the predictions with different PDF sets differ within their
respective uncertainty bands~\cite{Lohwasser:1265829,Alekhin:2011sk},
they follow the same global trend, rising with $\eta_\mu$. The
measured asymmetry agrees with this expectation. As demonstrated
graphically in Figure~\ref{fig:asym}, the data are roughly compatible
with all the predictions with different PDF sets, though some are
slightly preferred to others.  A $\chi^{2}$-comparison using the
measurement uncertainty and the central value of the PDF predictions
yields values per number of degrees of freedom of $9.16/11$ for the
CTEQ 6.6 PDF sets, $35.81/11$ for the HERA 1.0 PDF sets and $27.31/11$
for the MSTW 2008 PDF sets.

In summary, this letter reports a measurement of the $\Wboson$ charge
asymmetry in $pp$ collisions at $\sqrt{s} = 7\TeV$ performed in the
$\Wboson \rightarrow \mu \nu$ decay mode using $31~\ipb$ of data
recorded with the ATLAS detector at the LHC. Until the start of the
LHC, it has not been kinematically possible to precisely measure the
valence quark distributions and in particular to constrain the ratio
of $u/d$ quarks below $x\lesssim0.05$~as assessed by
\cite{Martin:2009iq}. Whereas none of the predictions with different
PDF sets are inconsistent with these data, the predictions are not
fully consistent with each other since they are all phenomenological
extrapolations in $x$. The input of the data presented here is
therefore expected to contribute to the determination of the next
generation of PDF sets, helping reduce PDF uncertainties, particularly
the shapes of the valence quark distributions in the low-$x$ region.

\renewcommand{\arraystretch}{1.3}
\begin{table*}[tbph]
\centering
\begin{tabular}{r|cccc}
\hline
\hline
 & Data & MSTW 2008 & CTEQ 6.6 & HERA 1.0\\
\hline
$0.00<|\eta_\mu|<0.21$ & $0.147 \pm 0.011 \pm 0.017$& $0.142_{-0.014}^{+0.006}$& $0.164_{-0.007}^{+0.006}$& $0.163\pm 0.007$\\
$0.21<|\eta_\mu|<0.42$ & $0.149 \pm 0.010 \pm 0.012$& $0.147_{-0.014}^{+0.007}$& $0.168_{-0.007}^{+0.006}$& $0.167\pm 0.007$\\
$0.42<|\eta_\mu|<0.63$ & $0.157 \pm 0.010 \pm 0.012$& $0.151_{-0.013}^{+0.007}$& $0.173_{-0.007}^{+0.006}$& $0.169\pm 0.007$\\
$0.63<|\eta_\mu|<0.84$ & $0.184 \pm 0.010 \pm 0.015$& $0.163_{-0.012}^{+0.008}$& $0.186_{-0.008}^{+0.007}$& $0.179_{-0.007}^{+0.008}$\\
$0.84<|\eta_\mu|<1.05$ & $0.186 \pm 0.011 \pm 0.017$& $0.176_{-0.012}^{+0.009}$& $0.198_{-0.008}^{+0.007}$& $0.188\pm 0.008$\\
$1.05<|\eta_\mu|<1.37$ & $0.239 \pm 0.008 \pm 0.011$& $0.197\pm 0.010$         & $0.219_{-0.010}^{+0.008}$& $0.203_{-0.008}^{+0.009}$\\
$1.37<|\eta_\mu|<1.52$ & $0.249 \pm 0.011 \pm 0.010$& $0.215_{-0.010}^{+0.011}$& $0.237_{-0.010}^{+0.009}$& $0.214\pm 0.009$\\
$1.52<|\eta_\mu|<1.74$ & $0.269 \pm 0.009 \pm 0.010$& $0.230_{-0.010}^{+0.012}$& $0.251_{-0.011}^{+0.009}$& $0.224\pm 0.009$\\
$1.74<|\eta_\mu|<1.95$ & $0.272 \pm 0.009 \pm 0.010$& $0.251_{-0.009}^{+0.013}$& $0.270_{-0.011}^{+0.010}$& $0.239_{-0.009}^{+0.010}$\\
$1.95<|\eta_\mu|<2.18$ & $0.277 \pm 0.009 \pm 0.012$& $0.266_{-0.010}^{+0.014}$& $0.284_{-0.011}^{+0.010}$& $0.251_{-0.010}^{+0.009}$\\
$2.18<|\eta_\mu|<2.40$ & $0.273 \pm 0.010 \pm 0.012$& $0.272_{-0.011}^{+0.015}$& $0.288_{-0.010}^{+0.009}$& $0.255_{-0.010}^{+0.009}$\\
\hline
\hline
\end{tabular}
\caption{The muon charge asymmetry from \Wboson-boson decays in bins
  of absolute pseudorapidity. The data measurements are listed with statistical
  and systematic uncertainties respectively. Predicted asymmetries of
  the MSTW 2008, CTEQ 6.6, and HERA 1.0 PDF sets are shown for
  comparison.}
\label{tab:asymvals}
\end{table*}

\section{Acknowledgements}

We wish to thank CERN for the efficient commissioning and operation of
the LHC during this initial high-energy data-taking period as well as
the support staff from our institutions without whom ATLAS could not
be operated efficiently.

We acknowledge the support of ANPCyT, Argentina; YerPhI, Armenia; ARC,
Australia; BMWF, Austria; ANAS, Azerbaijan; SSTC, Belarus; CNPq and
FAPESP, Brazil; NSERC, NRC and CFI, Canada; CERN; CONICYT, Chile; CAS,
MOST and NSFC, China; COLCIENCIAS, Colombia; MSMT CR, MPO CR and VSC
CR, Czech Republic; DNRF, DNSRC and Lundbeck Foundation, Denmark;
ARTEMIS, European Union; IN2P3-CNRS,\\ CEA-DSM/IRFU, France; GNAS,
Georgia; BMBF, DFG, HGF, MPG and AvH Foundation, Germany; GSRT,
Greece; ISF, MINERVA, GIF, DIP and Benoziyo Center, Israel; INFN,
Italy; MEXT and JSPS, Japan; CNRST, Morocco; FOM and NWO, Netherlands;
RCN, Norway; MNiSW, Poland; GRICES and FCT, Portugal; MERYS (MECTS),
Romania; MES of Russia and ROSATOM, Russian Federation; JINR; MSTD,
Serbia; MSSR, Slovakia; ARRS and MVZT, Slovenia; DST/NRF, South
Africa; MICINN, Spain; SRC and Wallenberg Foundation, Sweden; SER,
SNSF and Cantons of Bern and Geneva, Switzerland; NSC, Taiwan; TAEK,
Turkey; STFC, the Royal Society and Leverhulme Trust, United Kingdom;
DOE and NSF, United States of America.

The crucial computing support from all WLCG partners is acknowledged
gratefully, in particular from CERN and the ATLAS Tier-1 facilities at
TRIUMF (Canada), NDGF (Denmark, Norway, Sweden), CC-IN2P3 (France),
KIT/GridKA (Germany), INFN-CNAF (Italy), NL-T1\\ (Netherlands), PIC
(Spain), ASGC (Taiwan), RAL (UK) and BNL (USA) and in the Tier-2
facilities worldwide.

\bibliographystyle{model1-num-names}
\bibliography{WAsymMuonPaper}

\begin{thebibliography}{42}
\expandafter\ifx\csname natexlab\endcsname\relax\def\natexlab#1{#1}\fi
\providecommand{\bibinfo}[2]{#2}
\ifx\xfnm\relax \def\xfnm[#1]{\unskip,\space#1}\fi
\bibitem[{Berger et~al.(1989)Berger, Halzen, Kim, and
  Willenbrock}]{Berger:1988tu}
\bibinfo{author}{E.~L. Berger}, \bibinfo{author}{F.~Halzen},
  \bibinfo{author}{C.~S. Kim}, \bibinfo{author}{S.~Willenbrock},
\newblock \bibinfo{title}{{Weak boson production at Tevatron energies}},
\newblock \bibinfo{journal}{Phys. Rev.} \bibinfo{volume}{D40}
  (\bibinfo{year}{1989}) \bibinfo{pages}{83}.
\bibitem[{{ZEUS Collaboration}(2009{\natexlab{a}})}]{Chekanov:2008aa}
\bibinfo{author}{{ZEUS Collaboration}},
\newblock \bibinfo{title}{{Measurement of charged current deep inelastic
  scattering cross sections with a longitudinally polarised electron beam at
  HERA}},
\newblock \bibinfo{journal}{Eur. Phys. J.} \bibinfo{volume}{C61}
  (\bibinfo{year}{2009}{\natexlab{a}}) \bibinfo{pages}{223--235}.
\bibitem[{{ZEUS Collaboration}(2009{\natexlab{b}})}]{Chekanov:2009gm}
\bibinfo{author}{{ZEUS Collaboration}},
\newblock \bibinfo{title}{{Measurement of high-$Q^2$ neutral current deep
  inelastic $e^-p$ scattering cross sections with a longitudinally polarised
  electron beam at HERA}},
\newblock \bibinfo{journal}{Eur. Phys. J.} \bibinfo{volume}{C62}
  (\bibinfo{year}{2009}{\natexlab{b}}) \bibinfo{pages}{625--658}.
\bibitem[{{H1 Collaboration}(2003)}]{Adloff:2003uh}
\bibinfo{author}{{H1 Collaboration}},
\newblock \bibinfo{title}{{Measurement and QCD analysis of neutral and charged
  current cross sections at HERA}},
\newblock \bibinfo{journal}{Eur. Phys. J.} \bibinfo{volume}{C30}
  (\bibinfo{year}{2003}) \bibinfo{pages}{1--32}.
\bibitem[{{H1 and ZEUS Collaborations}(2010)}]{:2009wt}
\bibinfo{author}{{H1 and ZEUS Collaborations}},
\newblock \bibinfo{title}{{Combined Measurement and QCD Analysis of the
  Inclusive $ep$ Scattering Cross Sections at HERA}},
\newblock \bibinfo{journal}{JHEP} \bibinfo{volume}{01} (\bibinfo{year}{2010})
  \bibinfo{pages}{109}.
\bibitem[{Nakamura et~al.(2010)}]{Nakamura:2010zzi}
\bibinfo{author}{K.~Nakamura}, et~al.,
\newblock \bibinfo{title}{{Review of particle physics}},
\newblock \bibinfo{journal}{J. Phys. G} \bibinfo{volume}{37}
  (\bibinfo{year}{2010}) \bibinfo{pages}{075021}.
\bibitem[{Evans and Bryant(2008)}]{LHC:2008}
\bibinfo{author}{L.~Evans}, \bibinfo{author}{P.~Bryant},
\newblock \bibinfo{title}{{LHC Machine}},
\newblock \bibinfo{journal}{JINST} \bibinfo{volume}{3} (\bibinfo{year}{2008})
  \bibinfo{pages}{S08001}.
\bibitem[{Alekhin et~al.(2005)}]{Alekhin:2005dx}
\bibinfo{author}{S.~Alekhin}, et~al.,
\newblock \bibinfo{title}{{HERA and the LHC - A workshop on the implications of
  HERA for LHC physics: Proceedings Part A}}  (\bibinfo{year}{2005}).
  \bibinfo{note}{{hep-ph/0601012,
  http://www.desy.de/\~heralhc/proceedings/proceedings.html}}.
\bibitem[{{ATLAS Collaboration}(2010)}]{wzpaper}
\bibinfo{author}{{ATLAS Collaboration}},
\newblock \bibinfo{title}{{Measurement of the $W\to l \nu$ and $Z/\gamma^{*}\to
  ll$ production cross sections in proton-proton collisions at $\sqrt{s} =
  7~\TeV$ with the ATLAS detector}},
\newblock \bibinfo{journal}{JHEP} \bibinfo{volume}{12} (\bibinfo{year}{2010})
  \bibinfo{pages}{001}.
\bibitem[{{CMS Collaboration}(2011)}]{Khachatryan:2010xn}
\bibinfo{author}{{CMS Collaboration}},
\newblock \bibinfo{title}{{Measurements of Inclusive $W$ and $Z$ Cross Sections
  in $pp$ Collisions at $\sqrt{s}=7~\TeV$}},
\newblock \bibinfo{journal}{JHEP} \bibinfo{volume}{01} (\bibinfo{year}{2011})
  \bibinfo{pages}{001}.
\bibitem[{Lohwasser et~al.(2010)Lohwasser, Ferrando, and
  Issever}]{Lohwasser:2010sp}
\bibinfo{author}{K.~Lohwasser}, \bibinfo{author}{J.~Ferrando},
  \bibinfo{author}{C.~Issever},
\newblock \bibinfo{title}{{On direct measurement of the $W$ production charge
  asymmetry at the LHC}},
\newblock \bibinfo{journal}{JHEP} \bibinfo{volume}{09} (\bibinfo{year}{2010})
  \bibinfo{pages}{079}.
\bibitem[{{CDF Collaboration}(1998)}]{Abe:1998rv}
\bibinfo{author}{{CDF Collaboration}},
\newblock \bibinfo{title}{{Measurement of the lepton charge asymmetry in $W$
  boson decays produced in $p \bar{p}$ collisions}},
\newblock \bibinfo{journal}{Phys. Rev. Lett.} \bibinfo{volume}{81}
  (\bibinfo{year}{1998}) \bibinfo{pages}{5754--5759}.
\bibitem[{{CDF Collaboration}(2005)}]{Acosta:2005ud}
\bibinfo{author}{{CDF Collaboration}},
\newblock \bibinfo{title}{{Measurement of the forward-backward charge asymmetry
  from $W \to e \nu$ production in $p\bar{p}$ collisions at $\sqrt{s} = 1.96$
  TeV}},
\newblock \bibinfo{journal}{Phys. Rev.} \bibinfo{volume}{D71}
  (\bibinfo{year}{2005}) \bibinfo{pages}{051104}.
\bibitem[{{D\O~Collaboration}(2008{\natexlab{a}})}]{Abazov:2007pm}
\bibinfo{author}{{D\O~Collaboration}},
\newblock \bibinfo{title}{{Measurement of the muon charge asymmetry from $W$
  boson decays}},
\newblock \bibinfo{journal}{Phys. Rev.} \bibinfo{volume}{D77}
  (\bibinfo{year}{2008}{\natexlab{a}}) \bibinfo{pages}{011106}.
\bibitem[{{D\O~Collaboration}(2008{\natexlab{b}})}]{Abazov:2008qv}
\bibinfo{author}{{D\O~Collaboration}},
\newblock \bibinfo{title}{{Measurement of the electron charge asymmetry in $p
  \bar{p} \to W + X \to e \nu + X$ events at $\sqrt{s}$ = $1.96\,$TeV}},
\newblock \bibinfo{journal}{Phys. Rev. Lett.} \bibinfo{volume}{101}
  (\bibinfo{year}{2008}{\natexlab{b}}) \bibinfo{pages}{211801}.
\bibitem[{{CDF Collaboration}(2009)}]{Aaltonen:2009ta}
\bibinfo{author}{{CDF Collaboration}},
\newblock \bibinfo{title}{{Direct Measurement of the $W$ Production Charge
  Asymmetry in $p\bar{p}$ Collisions at $\sqrt{s} = 1.96$ TeV}},
\newblock \bibinfo{journal}{Phys. Rev. Lett.} \bibinfo{volume}{102}
  (\bibinfo{year}{2009}) \bibinfo{pages}{181801}.
\bibitem[{Martin et~al.(2009)Martin, Stirling, Thorne, and
  Watt}]{Martin:2009iq}
\bibinfo{author}{A.~D. Martin}, \bibinfo{author}{W.~J. Stirling},
  \bibinfo{author}{R.~S. Thorne}, \bibinfo{author}{G.~Watt},
\newblock \bibinfo{title}{{Parton distributions for the LHC}},
\newblock \bibinfo{journal}{Eur. Phys. J.} \bibinfo{volume}{C63}
  (\bibinfo{year}{2009}) \bibinfo{pages}{189--285}.
\bibitem[{Pumplin et~al.(2002)}]{Pumplin:2002vw}
\bibinfo{author}{J.~Pumplin}, et~al.,
\newblock \bibinfo{title}{{New generation of parton distributions with
  uncertainties from global QCD analysis}},
\newblock \bibinfo{journal}{JHEP} \bibinfo{volume}{07} (\bibinfo{year}{2002})
  \bibinfo{pages}{012}.
\bibitem[{{ATLAS Collaboration}(2008)}]{AtlasDetector}
\bibinfo{author}{{ATLAS Collaboration}},
\newblock \bibinfo{title}{{The ATLAS Experiment at the CERN Large Hadron
  Collider}},
\newblock \bibinfo{journal}{JINST} \bibinfo{volume}{3} (\bibinfo{year}{2008})
  \bibinfo{pages}{S08003}.
\bibitem[{{ATLAS Collaboration}(2009)}]{CSCbook}
\bibinfo{author}{{ATLAS Collaboration}},
\newblock \bibinfo{title}{{Expected Performance of the ATLAS Experiment -
  Detector, Trigger and Physics}},
\newblock \bibinfo{journal}{arXiv:0901.0512 [hep-ex]}  (\bibinfo{year}{2009}).
\bibitem[{Sjostrand et~al.(2006)Sjostrand, Mrenna, and Skands}]{pythia}
\bibinfo{author}{T.~Sjostrand}, \bibinfo{author}{S.~Mrenna},
  \bibinfo{author}{P.~Skands},
\newblock \bibinfo{title}{Pythia 6.4 physics and manual},
\newblock \bibinfo{journal}{JHEP} \bibinfo{volume}{05} (\bibinfo{year}{2006})
  \bibinfo{pages}{026}.
\bibitem[{Sherstnev and Thorne(2008)}]{mrst}
\bibinfo{author}{A.~Sherstnev}, \bibinfo{author}{R.~S. Thorne},
\newblock \bibinfo{title}{{Parton Distributions for LO Generators}},
\newblock \bibinfo{journal}{Eur. Phys. J.} \bibinfo{volume}{C55}
  (\bibinfo{year}{2008}) \bibinfo{pages}{553}.
\bibitem[{Frixione et~al.(2007)Frixione, Nason, and Oleari}]{Powheg}
\bibinfo{author}{S.~Frixione}, \bibinfo{author}{P.~Nason},
  \bibinfo{author}{C.~Oleari},
\newblock \bibinfo{title}{{Matching NLO QCD computations with parton shower
  simulations:the POWHEG method}},
\newblock \bibinfo{journal}{JHEP} \bibinfo{volume}{11} (\bibinfo{year}{2007})
  \bibinfo{pages}{070}.
\bibitem[{Nadolsky et~al.(2008)}]{CTEQ6.6M}
\bibinfo{author}{P.~M. Nadolsky}, et~al.,
\newblock \bibinfo{title}{{Implications of CTEQ global analysis for collider
  observables}},
\newblock \bibinfo{journal}{Phys. Rev.} \bibinfo{volume}{D78}
  (\bibinfo{year}{2008}) \bibinfo{pages}{013004}.
\bibitem[{Golonka and Was(2006)}]{Photos}
\bibinfo{author}{P.~Golonka}, \bibinfo{author}{Z.~Was},
\newblock \bibinfo{title}{{PHOTOS Monte Carlo: a precision tool for QED
  corrections in Z and W decays}},
\newblock \bibinfo{journal}{Eur. Phys. J.} \bibinfo{volume}{C45}
  (\bibinfo{year}{2006}) \bibinfo{pages}{97}.
\bibitem[{Davidson et~al.(2010)}]{Tauola}
\bibinfo{author}{N.~Davidson}, et~al.,
\newblock \bibinfo{title}{{Universal interface of TAUOLA technical and physics
  documentation}},
\newblock \bibinfo{journal}{{arXiv:1002.0543 [hep-ph]}}
  (\bibinfo{year}{2010}).
\bibitem[{{ATLAS Collaboration}(2010)}]{MC09tune}
\bibinfo{author}{{ATLAS Collaboration}},
\newblock \bibinfo{title}{{ATLAS Monte Carlo tunes for MC09}},
\newblock \bibinfo{journal}{ATLAS-PHYS-PUB-2010-002}  (\bibinfo{year}{2010}).
  \bibinfo{note}{{http://cdsweb.cern.ch/record/1247375}}.
\bibitem[{Agostinelli et~al.(2003)}]{geant4}
\bibinfo{author}{S.~Agostinelli}, et~al.,
\newblock \bibinfo{title}{Geant4: A simulation toolkit},
\newblock \bibinfo{journal}{Nucl. Instrum. Meth.} \bibinfo{volume}{A506}
  (\bibinfo{year}{2003}) \bibinfo{pages}{250}.
\bibitem[{{ATLAS Collaboration}(2010)}]{AtlasSimulation}
\bibinfo{author}{{ATLAS Collaboration}},
\newblock \bibinfo{title}{{The ATLAS Simulation Infrastructure}},
\newblock \bibinfo{journal}{Eur. Phys. J. C} \bibinfo{volume}{70}
  (\bibinfo{year}{2010}) \bibinfo{pages}{787}.
\bibitem[{Anastasiou et~al.(2004)Anastasiou, Dixon, Melnikov, and
  Petriello}]{FEWZ}
\bibinfo{author}{C.~Anastasiou}, \bibinfo{author}{L.~Dixon},
  \bibinfo{author}{K.~Melnikov}, \bibinfo{author}{F.~Petriello},
\newblock \bibinfo{title}{{High-precision QCD at hadron colliders: electroweak
  gauge boson rapidity distributions at NNLO}},
\newblock \bibinfo{journal}{Phys. Rev.} \bibinfo{volume}{D69}
  (\bibinfo{year}{2004}) \bibinfo{pages}{094008}.
\bibitem[{Martin et~al.(2009)Martin, Stirling, Thorne, and Watt}]{MSTW2008}
\bibinfo{author}{A.~D. Martin}, \bibinfo{author}{W.~J. Stirling},
  \bibinfo{author}{R.~S. Thorne}, \bibinfo{author}{G.~Watt},
\newblock \bibinfo{title}{{Parton distributions for the LHC}},
\newblock \bibinfo{journal}{Eur. Phys. J.} \bibinfo{volume}{C63}
  (\bibinfo{year}{2009}) \bibinfo{pages}{189}.
\bibitem[{Bonciani et~al.(1998)Bonciani, Catani, Mangano, and
  Nason}]{Bonciani:1998vc}
\bibinfo{author}{R.~Bonciani}, \bibinfo{author}{S.~Catani},
  \bibinfo{author}{M.~L. Mangano}, \bibinfo{author}{P.~Nason},
\newblock \bibinfo{title}{{NLL resummation of the heavy-quark hadroproduction
  cross-section}},
\newblock \bibinfo{journal}{Nucl. Phys.} \bibinfo{volume}{B529}
  (\bibinfo{year}{1998}) \bibinfo{pages}{424}.
\bibitem[{{ATLAS Collaboration}(2010{\natexlab{a}})}]{wjets}
\bibinfo{author}{{ATLAS Collaboration}},
\newblock \bibinfo{title}{{Measurement of the production cross section for
  $W$-bosons in association with jets in $pp$ collisions at $\sqrt{s}=7$~TeV
  with the ATLAS detector}},
\newblock \bibinfo{journal}{arXiv:1012.5382 [hep-ex]}
  (\bibinfo{year}{2010}{\natexlab{a}}). \bibinfo{note}{{Submitted to Phys.
  Lett. B}}.
\bibitem[{{ATLAS Collaboration}(2010{\natexlab{b}})}]{CLEANING}
\bibinfo{author}{{ATLAS Collaboration}},
\newblock \bibinfo{title}{Data-quality requirements and event cleaning for jets
  and missing transverse energy reconstruction with the atlas detector in
  proton-proton collisions at a center-of-mass energy of $\sqrt{s}=7\,${TeV}},
\newblock \bibinfo{journal}{{\scshape ATLAS} conference note: {\scshape
  ATLAS-CONF-2010-038}}  (\bibinfo{year}{2010}{\natexlab{b}}).
  \bibinfo{note}{{http://cdsweb.cern.ch/record/1277678}}.
\bibitem[{{ATLAS Collaboration}(2010{\natexlab{c}})}]{muperfcosmics}
\bibinfo{author}{{ATLAS Collaboration}},
\newblock \bibinfo{title}{{Commissioning of the ATLAS Muon Spectrometer with
  Cosmic Rays}},
\newblock \bibinfo{journal}{Eur. Phys. J.} \bibinfo{volume}{C70}
  (\bibinfo{year}{2010}{\natexlab{c}}) \bibinfo{pages}{875--916}.
\bibitem[{{ATLAS Collaboration}(2011)}]{muonreso}
\bibinfo{author}{{ATLAS Collaboration}}, \bibinfo{title}{Muon momentum
  resolution in first pass reconstruction of pp collision data recorded by
  ATLAS in 2010}, \bibinfo{type}{Technical Report}
  \bibinfo{number}{ATL-COM-CONF-2011-003}, \bibinfo{year}{2011}.
\bibitem[{Moch and Uwer(2008)}]{Moch:2008qy}
\bibinfo{author}{S.~Moch}, \bibinfo{author}{P.~Uwer},
\newblock \bibinfo{title}{{Theoretical status and prospects for top-quark pair
  production at hadron colliders}},
\newblock \bibinfo{journal}{Phys. Rev.} \bibinfo{volume}{D78}
  (\bibinfo{year}{2008}) \bibinfo{pages}{034003}.
\bibitem[{Beneke et~al.(2010)Beneke, Czakon, Falgari, Mitov, and
  Schwinn}]{Beneke:2009ye}
\bibinfo{author}{M.~Beneke}, \bibinfo{author}{M.~Czakon},
  \bibinfo{author}{P.~Falgari}, \bibinfo{author}{A.~Mitov},
  \bibinfo{author}{C.~Schwinn},
\newblock \bibinfo{title}{{Threshold expansion of the $gg(q\overline{q})
  \rightarrow Q\overline{Q}+X$ cross section at
  $\mathcal{O}(\alpha_{s}^{4})$}},
\newblock \bibinfo{journal}{Phys. Lett.} \bibinfo{volume}{B690}
  (\bibinfo{year}{2010}) \bibinfo{pages}{483}.
\bibitem[{Frixione and Webber(2002)}]{Frixione:2002ik}
\bibinfo{author}{S.~Frixione}, \bibinfo{author}{B.~R. Webber},
\newblock \bibinfo{title}{{Matching NLO QCD computations and parton shower
  simulations}},
\newblock \bibinfo{journal}{JHEP} \bibinfo{volume}{06} (\bibinfo{year}{2002})
  \bibinfo{pages}{029}.
\bibitem[{Pumplin et~al.(2001)}]{Pumplin:2001ct}
\bibinfo{author}{J.~Pumplin}, et~al.,
\newblock \bibinfo{title}{{Uncertainties of predictions from parton
  distribution functions. 2. The Hessian method}},
\newblock \bibinfo{journal}{Phys. Rev.} \bibinfo{volume}{D65}
  (\bibinfo{year}{2001}) \bibinfo{pages}{014013}.
\bibitem[{Lohwasser(2010)}]{Lohwasser:1265829}
\bibinfo{author}{K.~Lohwasser}, \bibinfo{title}{The W Charge Asymmetry:
  Measurement of the Proton Structure with the ATLAS detector}, Ph.D. thesis,
  University of Oxford, \bibinfo{address}{Oxford, UK}, \bibinfo{year}{2010}.
  \bibinfo{note}{{CERN-THESIS-2010-069. http://cdsweb.cern.ch/record/1265829}}.
\bibitem[{Alekhin et~al.(2011)}]{Alekhin:2011sk}
\bibinfo{author}{S.~Alekhin}, et~al.,
\newblock \bibinfo{title}{{The PDF4LHC Working Group Interim Report}},
\newblock \bibinfo{journal}{arXiv:1101.0536 [hep-ph]}  (\bibinfo{year}{2011}).

\end{thebibliography}

\newpage
\begin{flushleft}
{\Large The ATLAS Collaboration}

\bigskip

G.~Aad$^{\rm 48}$,
B.~Abbott$^{\rm 111}$,
J.~Abdallah$^{\rm 11}$,
A.A.~Abdelalim$^{\rm 49}$,
A.~Abdesselam$^{\rm 118}$,
O.~Abdinov$^{\rm 10}$,
B.~Abi$^{\rm 112}$,
M.~Abolins$^{\rm 88}$,
H.~Abramowicz$^{\rm 153}$,
H.~Abreu$^{\rm 115}$
% ...
\bigskip

% ...
$^{10}$ Institute of Physics, Azerbaijan Academy of Sciences, Baku, Azerbaijan\\
$^{11}$ Institut de F\'isica d'Altes Energies and Universitat Aut\`onoma  de Barcelona and ICREA, Barcelona, Spain\\
% ...
$^{48}$ Fakult\"{a}t f\"{u}r Mathematik und Physik, Albert-Ludwigs-Universit\"{a}t, Freiburg i.Br., Germany\\
$^{49}$ Section de Physique, Universit\'e de Gen\`eve, Geneva, Switzerland\\
% ...
$^{88}$ Department of Physics and Astronomy, Michigan State University, East Lansing MI, United States of America\\
% ...
$^{111}$ Homer L. Dodge Department of Physics and Astronomy, University of Oklahoma, Norman OK, United States of America\\
$^{112}$ Department of Physics, Oklahoma State University, Stillwater OK, United States of America\\
% ...
$^{115}$ LAL, Univ. Paris-Sud and CNRS/IN2P3, Orsay, France\\
% ...
$^{118}$ Department of Physics, Oxford University, Oxford, United Kingdom\\
% ...
$^{153}$ Raymond and Beverly Sackler School of Physics and Astronomy, Tel Aviv University, Tel Aviv, Israel\\
% ...
\end{flushleft}

\end{document}